\documentclass{article}

\usepackage{PRIMEarxiv}

\usepackage[utf8]{inputenc} 
\usepackage[T1]{fontenc}    
\usepackage{hyperref}       
\usepackage{url}            
\usepackage{booktabs}       
\usepackage{amsfonts}       
\usepackage{nicefrac}       
\usepackage{microtype}      
\usepackage{lipsum}
\usepackage{fancyhdr}       
\usepackage{graphicx}       
\graphicspath{{media/}}     

\usepackage{natbib}
\setcitestyle{authoryear,open={(},close={)}} 

\title{SmartCS: Enabling the Creation of ML-Powered Computer Vision Mobile Apps for Citizen Science Applications without Coding
\thanks{This is the submitted manuscript of a paper that was accepted in the \textbf{Citizen Science: Theory and Practice} journal on \textbf{May 15, 2024}. This is not the final version.} 
}

\author{
  Fahim Hasan Khan \\
  University of California, Santa Cruz \\
  Santa Cruz, CA 95064 \\
  \texttt{fkhan4@ucsc.edu} \\
   \And
  Akila de Silva \\
  University of California, Santa Cruz \\
  Santa Cruz, CA 95064 \\
  \texttt{audesilv@ucsc.edu} \\
  \AND
  Gregory Dusek \\
  NOAA National Ocean Service \\
  Silver Spring, MD 20910 \\
  \texttt{gregory.dusek@noaa.gov} \\
  \And
  James Davis \\
  University of California, Santa Cruz \\
  Santa Cruz, CA 95064 \\
  \texttt{davisje@ucsc.edu} \\
  \And
  Alex Pang \\
  University of California, Santa Cruz \\
  Santa Cruz, CA 95064 \\
  \texttt{pang@soe.ucsc.edu} \\
}

\begin{document}
\maketitle

\begin{abstract}
It is undeniable that citizen science contributes to the advancement of various fields of study. There are now software tools that facilitate the development of citizen science apps. However, apps developed with these tools rely on individual human skills to correctly collect useful data. Machine learning (ML)-aided apps provide on-field guidance to citizen scientists on data collection tasks. However, these apps rely on server-side ML support, and therefore need a reliable internet connection. Furthermore, the development of citizen science apps with ML support requires a significant investment of time and money. For some projects, this barrier may preclude the use of citizen science effectively. We present a platform that democratizes citizen science by making it accessible to a much broader audience of both researchers and participants. The SmartCS platform allows one to create citizen science apps with ML support quickly and without coding skills. Apps developed using SmartCS have client-side ML support, making them usable in the field, even when there is no internet connection. The client-side ML helps educate users to better recognize the subjects, thereby enabling high-quality data collection. We present several citizen science apps created using SmartCS, some of which were conceived and created by high school students.
\end{abstract}


\section{Introduction}
Citizen science is a form of scientific research that involves the general public as participants. The participants are typically not trained scientists, but rather individuals who are interested in or concerned about a particular subject and want to contribute to scientific knowledge \citep{aristeidou2020online, haklay2021citizen, vohland2021science}. Citizen science projects often involve monitoring and collecting visual data, such as images or videos, of various subjects. Participants may also be involved in designing experiments, analyzing results, and solving problems related to the research project. In the rest of this article, the term “researchers” refers to those who conduct the research projects, and “participants” refers to those who contribute to research projects via a citizen science platform \citep{eitzel2017citizen}. Citizen science benefits both researchers and participants in terms of data collection and learning experience. For example, the iNaturalist app allows participants to collect data while learning about plant and animal species \citep{van2018inaturalist}. 

With emerging technologies and shifting paradigms, citizen science platforms are also becoming mobile, making it easier for participants to collect visual data \citep{newman2012future}. Mobile devices, such as smartphones and tablets, equipped with multiple cameras and advanced processing capabilities, can perform complex computer vision (CV) tasks using machine learning (ML) \citep{Qin_2019_ICCV}. ML-enhanced customized mobile application software or apps have the potential to significantly improve the effectiveness of citizen science projects.

While anyone can engage in citizen science projects, some basic skills are necessary for effective data collection. Often, participants need to accurately identify and label objects, a task that can be challenging for non-experts. Mobile apps with integrated machine learning (ML), capable of detecting objects of interest, can assist participants in efficiently collecting visual data and improving the data quality. This also opens the possibility of recruiting more volunteers by educating people about new topics and growing new interests among them \citep{roche2020citizen}.

Citizen science platforms such as Zooniverse \citep{simpson2014zooniverse}, SPOTTERON \citep{hummer2018don}, Anecdata \citep{disney2017next}, etc., provide the service to build citizen science apps for crowdsourced research projects \citep{liu2021citizen}. While these platforms offer standardized purpose-specific tools and features, ML guidance is largely unavailable. A few apps, such as iNaturalist, have ML guidance through cloud servers. However, the required connectivity may be unavailable in remote locations. Moreover, existing open source systems like iNaturalist and Zooniverse are not designed to integrate with client-side ML \citep{simpson2014zooniverse}. Creating a new app with ML guidance de novo for every similar citizen science project is not ideal. For instance, ‘Seek by iNaturalist’ was built from scratch to add client-side ML guidance for classifying plant and animal species, despite iNaturalist having an existing ML server \citep{van2018inaturalist}. 

This paper introduces an ML-integrated citizen science mobile app creation platform for faster app building and deployment without programming knowledge. Developing apps for citizen science involves considering many factors, such as design and technical build \citep{lemmens2021citizen}. Furthermore, the investment required for app design and development often spans from tens to hundreds of thousands of US dollars \citep{odenwald2019smartphone}. The cost of crafting apps, especially those with an extensive range of features can hinder innovation \citep{howard2022review}. With SmartCS, users can bypass the complexities inherent in app development. Our platform offers pre-built features and templates within a single framework. This facilitates rapid prototyping and faster deployments. In addition to helping participants capture better data, the apps created by this platform can serve as educational tools to increase engagement in a wide variety of citizen science projects. We validate our platform's usefulness to “researchers” by asking a group of non-programmers to create apps and measuring their success and comfort in doing so. We further validate our platform’s usefulness to the “participants” by comparing success at correctly identifying the subjects of data to be captured and through a survey of participant engagement.

\section{Motivation}
Here, we discuss the challenges participants face in research data collection for rip current detection \citep{philip2016detecting}. Rip currents are strong, seaward flowing currents that can occur on any beach with breaking waves, leading to an estimated 100 drownings a year in the US \citep{castelle2016rip,gensini2010examination}. To answer questions like: ‘Which beaches have rip currents?’, the researcher needs to gather and label imagery that contains rip currents from many different geographic locations. While an expert can visually spot rip currents, it can be challenging for non-experts \citep{brannstrom2015you}. However, various ML methods can detect rip currents, as demonstrated by recent works \citep{de2023ripviz, de2021automated, maryan2019machine}. Mobile apps empowered by ML can assist non-expert data collectors by visually showing them the detected rip currents using bounding boxes or similar visualization through the live camera feed.

The same concept as the example above applies to data collection in other fields, such as biological sciences, marine life, geomorphology, weather related phenomena, etc. The ML-empowered apps allow non-expert participants to learn and correctly detect the subjects through on-the-field assistance in these data collection scenarios. This is especially helpful for relatively hard-to-recognize or differentiate objects, such as rip currents.

\section{Related Works}
\subsection{Citizen Science Platforms}
We examined several popular citizen science app creation platforms enabling people-powered mobile app development \citep{liu2021citizen}. Earlier, we mentioned Zooniverse, which features projects from diverse domains \citep{barber2018zooniverse, simpson2014zooniverse}. One of the Zooniverse projects is OceanEYEs \citep{rakeeping}, which seeks volunteers to count and label fish, if there are any, in millions of unlabeled images collected from the ocean. Anecdata \citep{disney2018anecdata} is another free platform like Zooniverse, where both are primarily web portals with companion mobile apps. SPOTTERON \citep{liu2021citizen} is a similar platform that exclusively functions through mobile apps with a uniform, easily customizable graphical user interface (GUI) for various projects. However, the base systems of these general-purpose citizen science app creation platforms do not support complex operations like ML model integration. The Citizen Science Association \citep{Platform50:online} also maintains a list of major citizen science platforms, complete with a comparison table highlighting their most prominent features. ML-assisted data collection is not included in this comparison, as this feature is not common on these platforms \citep{, CyberTracker, CitSci, ispotnature, Epicollect5}.

\subsection{Citizen Science Apps with ML}
Many custom-built citizen science applications have ML capabilities. We previously mentioned that iNaturalist \citep{van2018inaturalist} has server-side ML capabilities and functions like a social network to connect nature observers. Leafsnap \citep{kumar2012leafsnap}, a mobile app for automatic plant species identification, is another example of a citizen science app that utilizes computer vision. Despite having many powerful features, Leafsnap's ML processing is performed on a cloud server, like iNaturalist, after the participants upload their images. Leafsnap's server-side ML processing took 5.4 seconds per image, which is not fast enough to produce real-time results \citep{kumar2012leafsnap}. Although modern high-end servers perform much faster ML operations, server-side ML is not feasible for many real-time applications, especially citizen science apps intended for use in remote locations. Wildme.org \citep{HomeWild65:online}, Fathomnet \citep{katija2022fathomnet,FathomNe88:online}, and many other web-only citizen science platforms also use server-side ML and have no mobile apps. Many existing applications were not designed as general-purpose citizen science apps, so while some are open source, developing a new app using any of them as a starting point would require significant programming expertise. \citep{khan2021authorin} presented a short paper discussing the integration of ML into citizen science projects. However, their approach required significant programming expertise to be effectively utilized at that time. A few apps, such as Pl@ntNet, incorporate client-side ML \citep{goeau2013pl,PlantNet}. Pl@ntNet is one of the most popular science apps and was also among the first to include client-side ML support, allowing it to work without an internet connection. However, as it is specifically designed for detecting plants, it cannot be used for other citizen science projects. However, similar to Seek by iNaturalist, it is specifically designed for detecting plants and cannot be used for other applications \citep{van2018inaturalist}.

\subsection{Mobile App Creation Platforms}
The development of mobile apps involves a challenging process that includes various phases such as platform selection, writing, debugging, optimizing code, creating user interfaces, simulation, testing, and support \citep{joorabchi2013real}. Few customizable mobile app creation and deployment platforms, such as App Movement \citep{garbett2016app}, provide generic templates for the community to build apps. Model-driven development (MDD) \citep{balagtas2008model} has been adopted for mobile app development to simplify the process, reducing technical complexity and costs significantly. For example, \cite{gharaat2021alba} proposed an MDD framework that enables novice app developers to model location-based apps by code transformation. Although MDD has improved the app creation process for developers, it does not support app customization and adding advanced features, such as ML, without writing code. \cite{li2015platform} proposed a platform to create mobile augmented reality apps without programming, but none of these systems support the integration of ML capabilities in the app.

Mobile applications are generally classified into three categories: native apps, hybrid apps, and mobile web apps \citep{huy2012evaluation, umuhoza2016model}. Native mobile apps are created specifically for a single platform (e.g., Google’s Android or Apple’s iOS) and leverage the hardware and software of that platform to enhance the user experience. Hybrid apps, on the other hand, run through web browsers after installing on devices like native apps and are typically created like webpages. Hybrid apps are more capable of streamlining the development process but are not as fast or reliable as native apps. Web apps are adaptive websites that change layout, outlook, and accessibility when accessed from a mobile device. Due to the technical complexity of integrating client-side ML models, native apps are the most feasible option for creating ML-powered apps. ML model training platforms such as Roboflow \citep{ciaglia2022roboflow}, Lobe.ai \citep{Lobe}, Ultralytics HUB \citep{yolov8}, etc., have basic app templates that can be used with the trained models. TensorFlow Lite \citep{tensorflow2015-whitepaper} provides similar simplified app templates. However, these templates are too basic for creating any real app without writing code and making substantial modifications. We explored various app-making platforms without coding available for mobile and PC platforms \citep{AppBuild5:online, AppsGeys63:online, AndromoM29:online, Nocodeap11:online}. However, these platforms primarily function as GUI makers with standard basic functionalities, such as text inputs and outputs, loading graphics, maps, calendars, websites, accessing system camera apps, etc. Integrating the complex process of loading ML models and providing computer vision functionalities is not available on any of these platforms.

\subsection{ML Models for Computer Vision on Mobile Devices}
Typically, mobile devices have limited computational resources and power, which are major constraints for running ML models on such devices. Many recent research projects have focused on creating optimal models for computer vision tasks, in terms of accuracy, speed, resource efficiency, scalability, robustness, and generalizability. They employ deep learning models, which can be categorized into two types based on their underlying structures: one-stage and two-stage. The trend discovered in the current research describes how highly efficient one-stage paradigms will prioritize the prediction speed in frames per second (FPS). In contrast, two-stage models strive to achieve the best per-frame accuracy by employing a filtering stage and predicting stage. Due to the limited computational resources of mobile devices, using two-stage models can be costly. As a solution, employing lightweight one-stage models for data collection and analysis on these devices can enable integration into a wider range of mobile applications that have computational constraints \citep{khan2021authorin}.

Modern vision tasks, such as object detection, image classification, semantic segmentation, etc., are mostly done using convolutional neural networks (CNN). As the previously discussed categorization is about all ML models in general, specifically for object detection, the two categories of CNN-based ML models to choose from are (1) region-based detectors and (2) single-shot detectors (SSD). The computational resource-intensive two-stage region-based detectors, such as Faster R-CNN \citep{ren2015faster}, have a region-proposal stage and a classifier stage. The limited computational power of today's mobile devices precludes their direct use. On the other hand, they have been utilized for real-time object detection via remote GPU servers \citep{Lee2017}. However, these server-dependent systems are only suitable for deployment in locations with network access. On the contrary, SSD attains object detection using a single-stage CNN \citep{Lee2017}. YOLO \citep{adarsh2020yolo}, EfficientDet \citep{tan2020efficientdet}, and SSDs, such as SSD MobileNet \cite{sandler2018mobilenetv2, chiu2020mobilenet}, are designed to perform in realtime sacrificing some accuracy \citep{huang2017speed}. Additionally, Sun et al. \citep{sun2020object} empirically showed that the SSD MobileNetv2 required the least amount of memory, thus making the SSD MobileNetv2 preferable for processing on mobile platforms. Similarly, variants of YOLOv8 \citep{yolov8}, EfficientNet \citep{tan2019efficientnet}, Inception \citep{xia2017inception}, and MobileNet \citep{chiu2020mobilenet} are optimized for real-time image classification on mobile devices. Likewise, YOLOv8-seg (Ultralytics 2023), U-Net MobileNetv2 \citep{siddique2021u}, MobileNetv2-DeepLab-v3 \citep{chen2017deeplab}, etc. provide real-time segmentation on portable devices. We select and use the ML models most suitable for computer vision mobile apps for citizen science applications.

\begin{figure}[ht]
    \centering
    \includegraphics[width=1\linewidth]{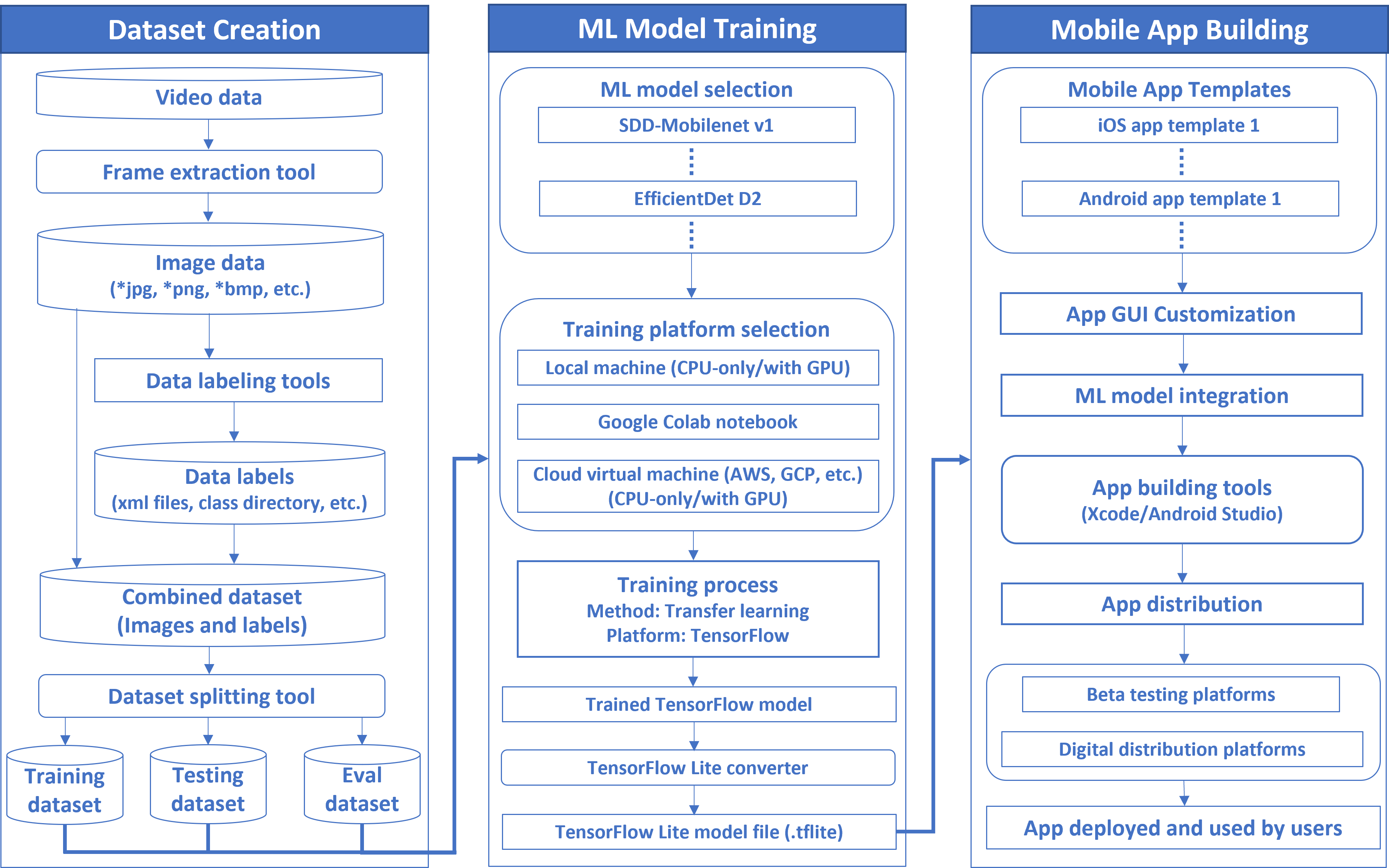}
    \caption{Overview and workflow of the components of our open-source citizen science app creation platform.}
    \label{fig:figure1}
\end{figure}

\section{System Design of the Platform}
Our citizen science app creation platform combines multiple technical components to build the overall system. The main parts of our proposed system are three steps: (1) Training dataset creation, (2) ML model training, and (3) Mobile app (iOS or Android) building. The workflow of our system is presented in Figure 1. Furthermore, our platform automates these steps.

The dataset consists of individual images or images extracted from videos to train ML models for computer vision. The type of labels needed for the image data depends on the type of task, such as object detection or image classification. The training images are labeled using bounding boxes around the objects to train a model for object detection. Labeling for image classification involves organizing the images into classes and labeling each class. Our system provides the required tools with instructions for formatting, labeling, and creating the training dataset.

Next, an ML model compatible with our system needs to be selected from a list. For each project, sufficient initial training data is needed to train a functioning model, which can be later improved via further training as more data are collected \citep{schwarz2018progress}. For a project with an adequate dataset, the researchers can immediately train a model and quickly deploy the app. However, the project team must create a minimum training dataset for a project without initial data. Our platform includes the option to add an "Expert mode (No ML)" to the apps. This feature allows volunteers to contribute to building up a dataset without relying on ML guidance. A back-end server  collects these data, which can then be curated. After quality control, feedback can be provided to volunteers to facilitate learning, as well as for retraining models to improve their accuracy. Our system provides built-in guidance to run the training process on a local machine or a cloud server.

\begin{figure}[ht]
    \centering
    \includegraphics[width=1\linewidth]{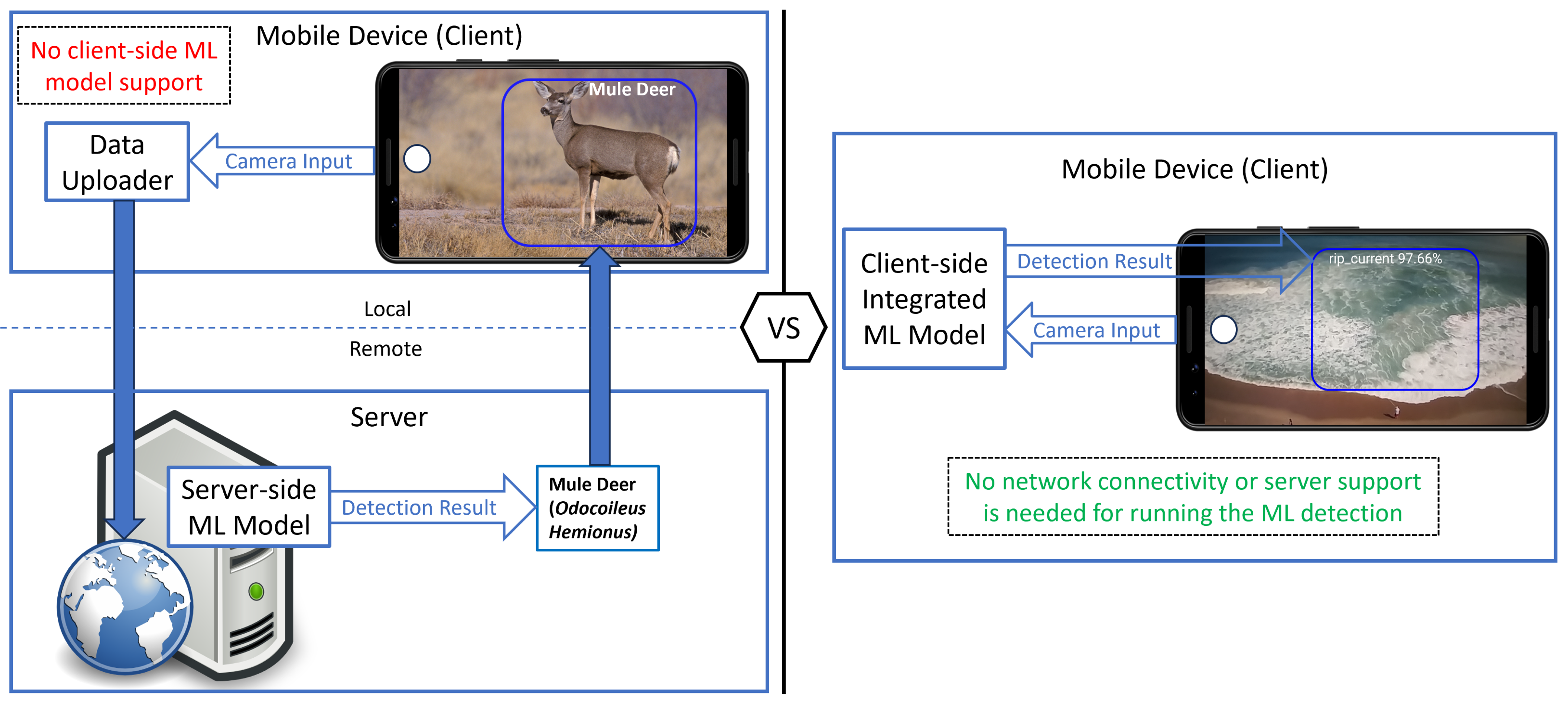}
    \caption{Server-side (left) vs client-side (right) machine learning (ML) models. We used client-side ML models in our implementation, which provided real-time object detection without any network connectivity or server-side processing requirements.}
    \label{fig:figure2}
\end{figure}

Finally, a template for the iOS or Android app needs to be selected from a collection of templates to fit the requirements of various citizen science projects. The template can be further customized to change GUI color, icon, logo, etc. Then, the app is built with the trained ML model integrated into it. The app’s primary visual data collection tool is like a camera app with a live view with image and video capture function, where the detection results are shown using visualizations, such as bounding boxes and text labels. These visualizations help participants identify and record data on objects of interest. The models operate on mobile devices using native computational resources, without any server support (Figure 2). A server is only necessary for uploading the collected data. Tools for data uploading, user guides, and tutorials are also included in the templates.

\section{Implementation}
We created a desktop and a web-based version of our app creation platform to make it more accessible and versatile. For both versions, the apps are created through the three simple guided steps. Our platform is implemented as an open-source tool, utilizing other open-source software and tools such as AnyLabeling, TensorFlow, PyTorch, and Colab Jupyter Notebook \citep{tensorflow2015-whitepaper,labelingtools}. An important advantage of open-source is that it enables users with programming skills to help improve the platform by writing code to update and create new features, creating new templates, etc.

\begin{figure}[ht]
    \centering
    \includegraphics[width=1\linewidth]{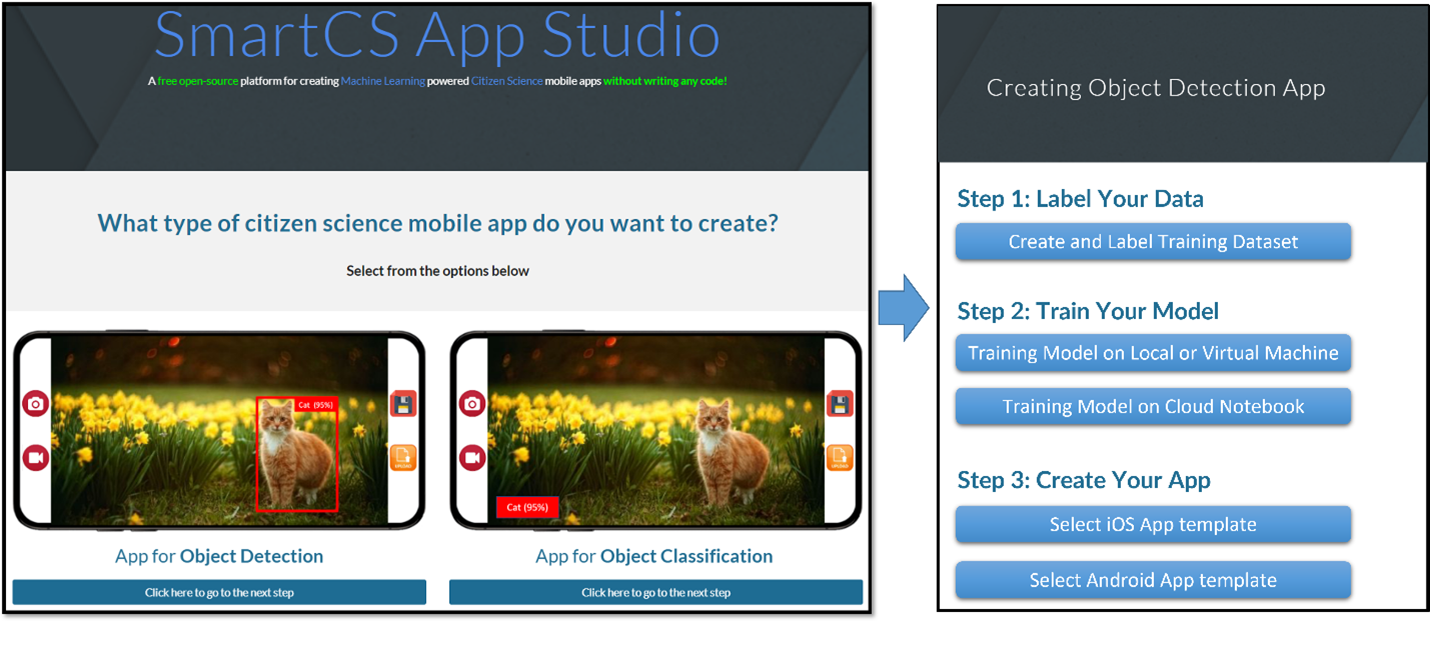}
    \caption{The web version of the platform was created for easy access and uses different feasible computational resources for different steps.}
    \label{fig:figure3}
\end{figure}

The desktop version is convenient for the user who wants to download it and run all the steps on a single machine capable of handling the ML training and app compilation. We created the desktop version to run on Windows, macOS, and Linux. The web-based version works as an online service, providing the user interface as a website, which can be accessed through any device that has a standard web browser (Figure 3). Another advantage of the web-based version is that different steps can run on different local, or cloud machines optimized for the specific step. For example, data labeling can be done on a laptop, as low computational resources are required. Then, the labeled data can be transferred to a GPU-optimized local or cloud machine for ML training. Because of this flexibility, the web-based version requires executing a few commands on the console and some file transfer operations. On the desktop version, the steps can be performed entirely by clicking some buttons and following the prompts, making it easy to be used by computer users with any skill level. Neither version requires programming skills or coding expertise.

\subsection{Dataset Creation:}
The object detection ML model training process expects still images as training data. We created a tool for extracting frames from the videos as still images if the available training data consists of video files. The frame extraction rate can be adjusted depending on the motion change rate of objects of interest. The data is labeled using free, open-source tools such as LabelImg \citep{labelingtools} or AnyLabeling (AI Curious 2023). Our platform also supports converting labels obtained using Amazon’s Mechanical Turk \citep{paolacci2010running}. Additionally, we a provide some custom tools that we created using Python for additional dataset formatting, such as image format conversion, annotation file processing, etc. The datasets for the classification tasks are created by simply putting the images for each class into separate directories, where each directory’s name works as the class names. Our system provides a tool to split the data into three sets, training, test, and evaluation, using a default ratio of 6:2:2 or user-defined splitting ratio \citep{reitermanova2010data}.

\subsection{ML Model Training:}
This next step allows the user to train the model on a local computer, a cloud virtual machine, or a cloud Python notebook. Python notebooks available via Google Colab provide access to free GPU resources. We use small and lightweight models optimized for mobile devices from the TensorFlow Lite library for training \citep{tensorflow2015-whitepaper}. Our platform provides the options to select the most appropriate model based on the information provided in Appendix B Table 1 and the planned usage of the app \cite{sanchez2020review}. Note that detection accuracy of the utilized ML models is considered reliable based on its Mean Average Precision (mAP) \citep{padilla2020survey}. The models are then custom trained using the dataset created in the previous step by the transfer learning process \citep{alsing2018mobile}. We provide tutorial documentation and video for model training.

While the training runs, a console shows the loss function value for the current training step. The lower the loss, the better a model has learned to detect the objects \citep{wang2022comprehensive}. The training needs to run until the model converges, which is indicated when the loss function value drops below a certain threshold \citep{allen2019convergence}. 
The training time depends on many factors such as the ML model, hardware (CPU only or GPU, amount of system memory, etc.) used for the training, and the training dataset (size, number of classes, etc.) \citep{lim2000comparison, jordan2015machine}. Further discussion about guidance on setting up these training parameters in the platform is included in Appendix B. ML model training for the case studies presented in this paper took about 2 to 3 days using inexpensive CPU-only machines, which can be done within a few hours using more expensive GPU machines.

\subsection{Mobile App Building:}
After training, the models are converted to TensorFlow Lite format, compatible with the platform's iOS and Android app templates. The app is built using the platform-specific build tool (Xcode for iOS or Android Studio for Android) and uploaded to digital distribution platforms for deployment. A paid developer account is required for Android and iOS to distribute the apps publicly through the official app distribution services. However, for initial app testing, it can be freely installed on a mobile device by connecting it to the USB cable to the computer where it was built. For easy deployment of apps on the Google Play Store and iOS App Store, we integrated Fastlane \citep{katz2018continuous} into our platform. Fastlane is an open-source platform that automates beta deployments and releases for mobile Android and iOS apps.

For user accounts, citizen science projects, and app management on our platform, we utilized Firebase Authentication \citep{moroney2017using}, an open-source user account management system. It facilitates managing user identities and authentication securely using various sign-in methods, including email and password, anonymous, and federated identity providers.

\section{Results}
This section presents a wide variety of citizen science apps created using SmartCS. The apps use ML models to aid users in collecting data for citizen science projects or as educational tools. Three apps, RipSnap, Seal vs. Sea Lion, and Vehicle Object Detection, were created for university research projects. In contrast, three other apps, Recycle This, TidalNow, and Sk.in, were created by high school students with no prior programming experience. Due to space constraints, we provide details on two of these example apps and briefly describe the other four use cases in Appendix A.

\begin{figure}[ht]
    \centering
    \includegraphics[width=1\linewidth]{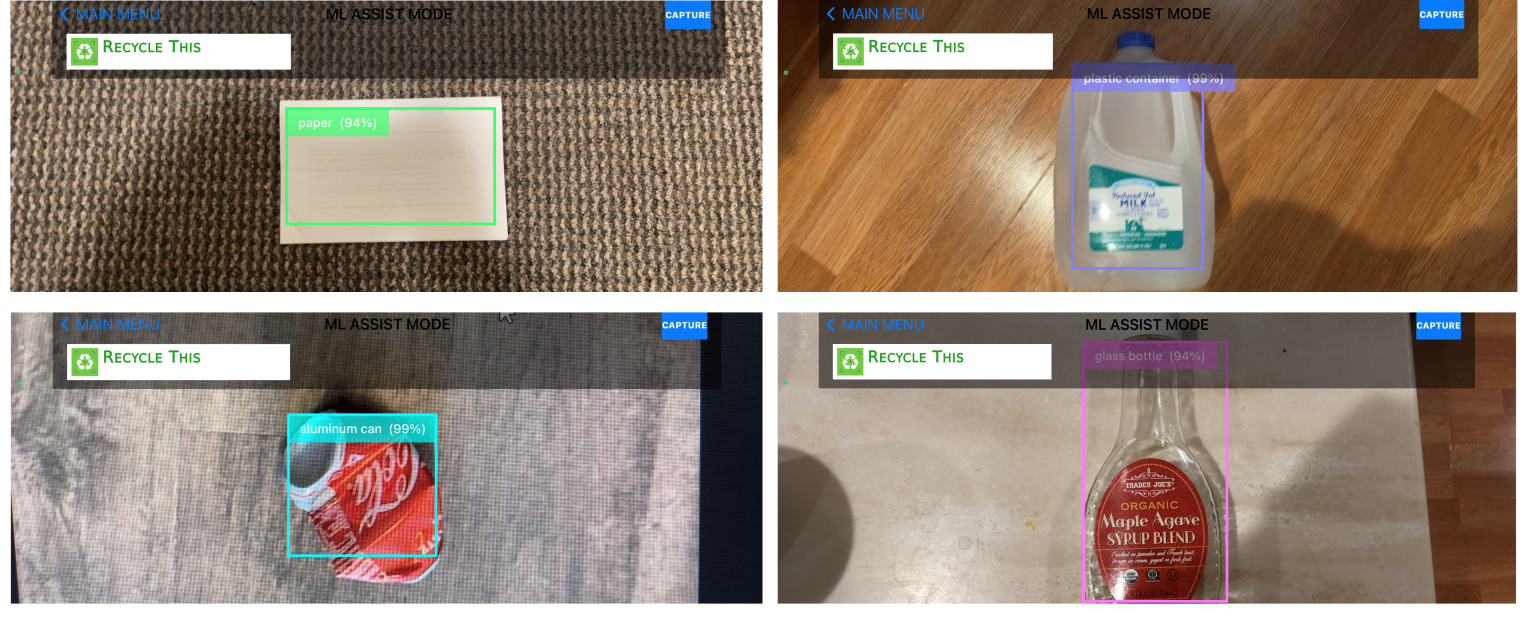}
    \caption{This figure illustrates the type of materials that the “Recycle This” app can detect papers, aluminum cans, plastic containers, and glass bottles.}
    \label{fig:figure4}
\end{figure}

\subsection{Use Case: Recycle This}
The practice of recycling is essential for the environment and the future of our planet \citep{al2009recycling}. Data collection on recyclable objects is necessary for optimizing recycling processes. However, there is a lack of clarity about what and how to recycle, with surveys showing that up to 62 percent of Americans lack recycling knowledge (Informa Markets, 2019). The mobile app Recycle This, created by a high school student, incorporates ML for real-time detection and collection of data on common household recyclables (Figure 4). The project focused on classifying objects like glass bottles, plastic containers, cardboard, paper, and aluminum cans. 2,500 training images were sourced from the public image datasets \citep{kaggle,imagescv,roboflow}. In addition to being a data collection tool, the app also acts as an educational tool by providing information and clarification on the recyclability of everyday waste objects. With widespread distribution, it can raise public awareness. SmartCS enabled the creation of the app without writing codes, and subsequent studies were published as conference papers \citep{yehrecycle}.

\begin{figure}[ht]
    \centering
    \includegraphics[width=1\linewidth]{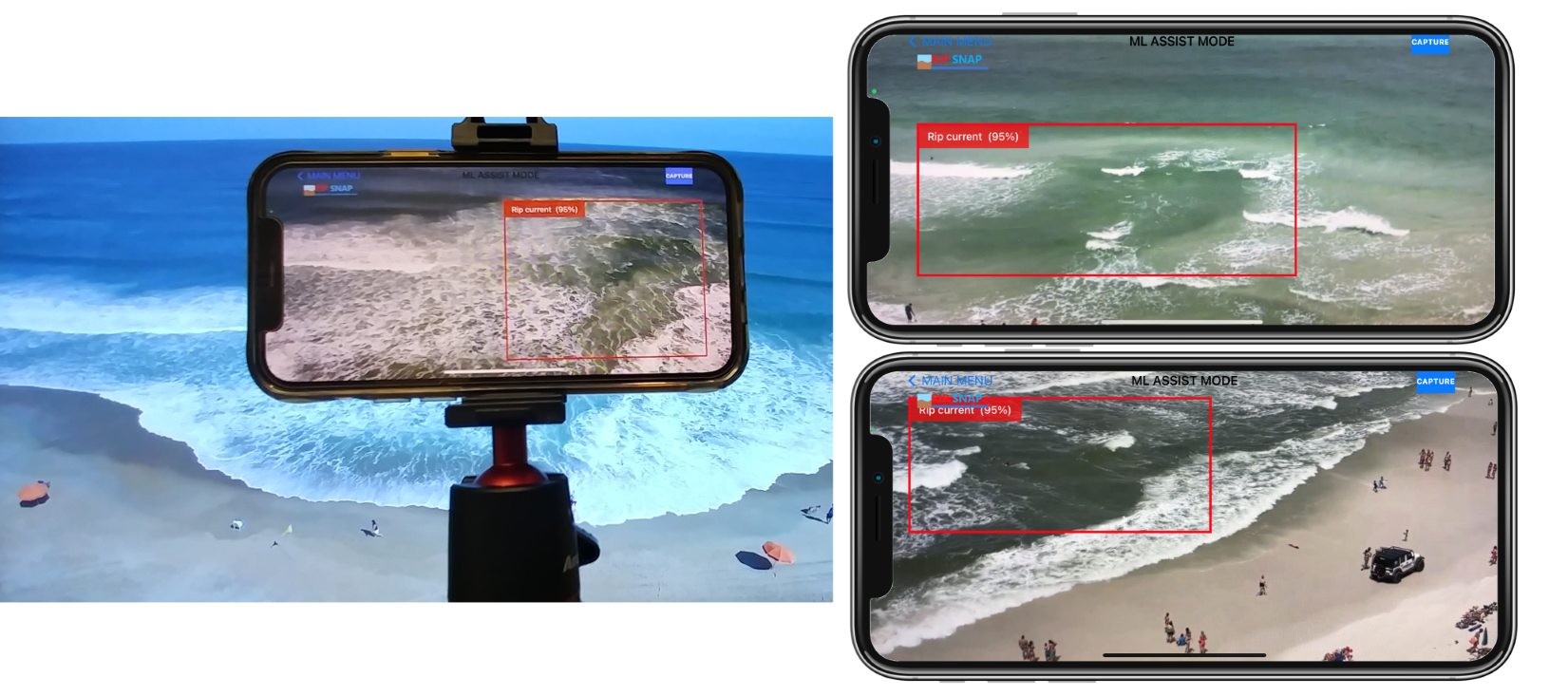}
    \caption{Appearance of the RipSnap app with examples of rip currents detected by the app. The location of the rip current is visualized using the red bounding box with the label and the confidence score of detection.}
    \label{fig:figure5}
\end{figure}

\subsection{Use Case: RipSnap}
The importance of rip current detection was discussed earlier. The citizen science app RipSnap is based on the idea of CoastSnap \citep{harley2018coastsnap}, where app users contribute snapshots of coastlines from fixed docking stations to study coastal erosion and other processes. RipSnap extends the idea of collecting ML-detected videos of rips with location metadata. Data collected through RipSnap can help validate a rip current forecast model \citep{dusek2013probabilistic}. The app’s primary aim is to enhance beach safety by educating users about the presence of rip currents (Figure 5). A lightweight ML model integrated into this app assists non-expert participants in identifying and collecting these valuable rip current data. The training dataset consists of 3,360 labeled images of two classes of rip currents, a combination of datasets from \citep{de2021automated}, and additional data collected using drones and beach monitoring webcams.

\begin{figure}[ht]
    \centering
    \includegraphics[width=0.75\linewidth]{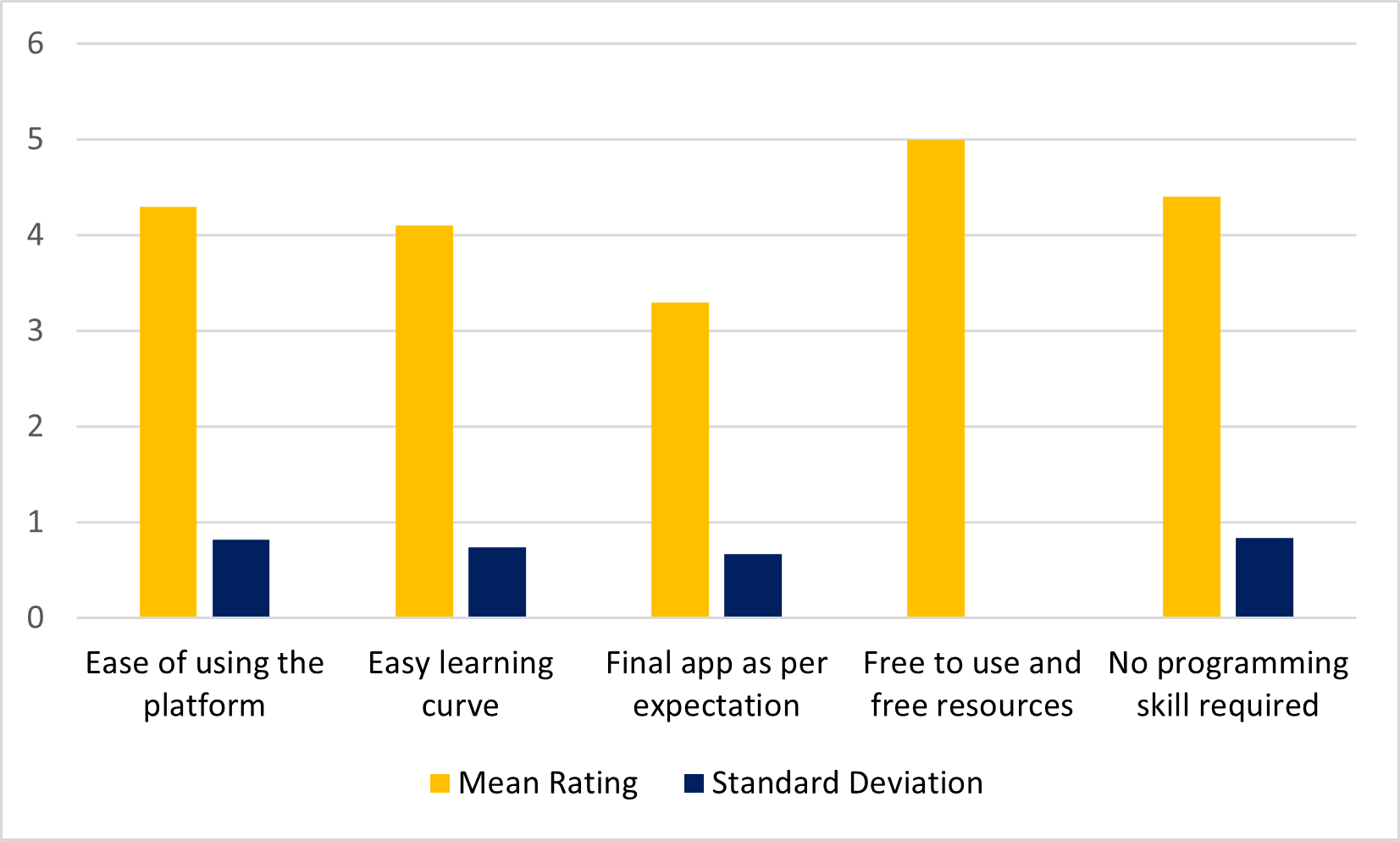}
    \caption{Summary of scores from the user study by app creators. Participants were asked to rate their experience of using the platform on a 5-point Likert scale from “Poor” (a score of 1) to “Excellent” (a score of 5).}
    \label{fig:figure6}
\end{figure}

\section{Feedback}
We collected user feedback from two user groups: (1) researchers who used the platform to create their apps, and (2) beta testers who used the created apps. The research questions and objectives are described before each study \citep{lazar2017research}. The main goals of these user studies were usability testing of the app creation platform and establishing the effectiveness of the created apps. The findings of the user studies are discussed in this section.

\subsection{User Study 1: App Creators }
Our research question for this user study is, “Will people without programming experience be able to create citizen science apps using SmartCS?” The objective is to collect feedback and evidence demonstrating that individuals without programming experience can successfully create citizen science apps using SmartCS.

This study gathered insights from 10 high school students, with 5 male and 5 female participants, all of whom did not have prior programming experience and were selected among science internship program participants by an independent committee. These students, ranging in age from 14 to 17 years, engaged as researchers to develop their own applications utilizing SmartCS. These students were tasked with individually creating apps, a challenge that all except one were able to successfully complete. The process of app creation took them between 1 to 2 weeks, a timeframe that varied depending on the complexity of data labeling and the time required for machine learning model training, which in turn was influenced by the number of available GPUs. More details about the apps are provided in Appendix A.  

For this study, we collected both quantitative and qualitative data from users. Participants' experiences and interactions with SmartCS were recorded using an online form, facilitating the collection of quantitative data through Likert scales. Additionally, the feedback form enabled the gathering of qualitative data through multiple open-ended responses. We also carefully observed participants' interactions with the system to gain deeper insights into their user experience. We observed that users could easily learn to use the platform by following the provided user guides and tutorials. Whenever users required additional help beyond what was available in the existing resources, we updated the guides and tutorials accordingly. Their main task was to learn how to navigate the platform's GUI. These observations helped us identify patterns and challenges within the user interface and interaction design that were not immediately apparent through quantitative feedback alone.

Figure 6 summarizes the quantitative user feedback on the app creation platform. Despite the small sample size, we can obtain some insights from the survey. The users were quite satisfied with the “Free to Use” aspect, as everyone gave it the highest rating. The “Final App as Per Expectation” feature has the lowest mean score, indicating the diversity of expectations among users. However, this may be due to the limited templates and customization options available in the current version, as suggested by the users' qualitative feedback regarding their unmet expectations. Users expressed the need for more application-specific features and customizations tailored to each app's unique purpose. However, providing such a high level of customization is challenging with a general-purpose tool, requiring a balance between ease of use and flexibility. This limitation can be partially addressed by providing more templates. The “No Programming Skill Required” question has the highest variability, which may be due to the different levels of computer exposure among the users. Overall, this feedback suggests that the platform is user-friendly and convenient for non-expert users.

\begin{figure}[ht]
    \centering
    \includegraphics[width=1\linewidth]{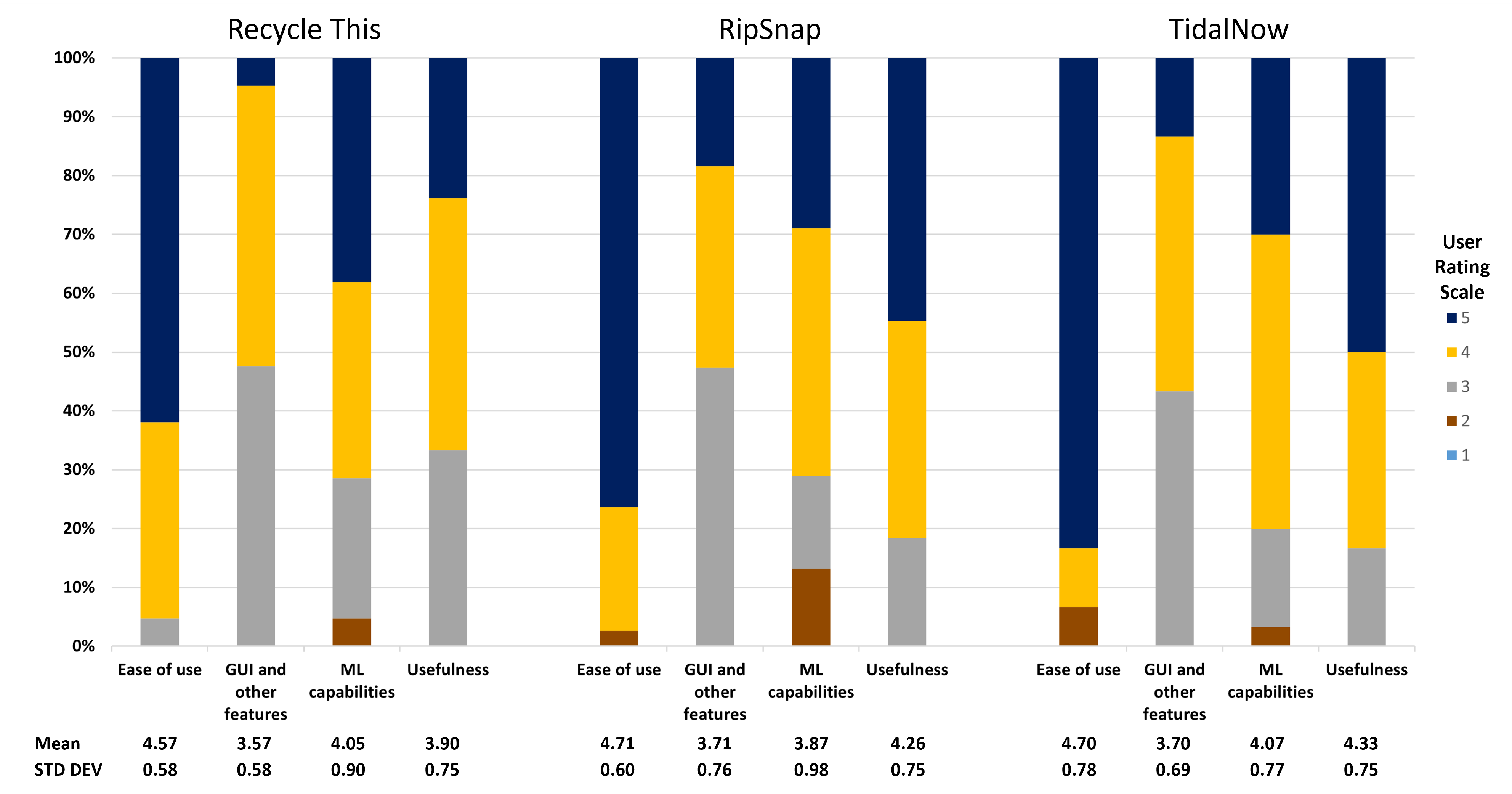}
    \caption{Summary of scores from the user study by users of three apps. Participants were asked to rate their experience of using the platform on a 5-point Likert scale from “Poor” (a score of 1) to “Excellent” (a score of 5).}
    \label{fig:figure7}
\end{figure}

\subsection{User Study 2: App Users }
The research question for user study 2 is, “Are the citizen science apps created using SmartCS useful and easy to use?” The objective is to collect feedback and evidence demonstrating that citizen science apps developed with SmartCS, featuring simple designs and ML guidance, are both useful and easy to use.
Three apps, RipSnap, Recycle This, and Tidal-Now, underwent user testing with 38, 21, and 30 testers, respectively. Participants provided feedback on their user experience through an online form (Figure 7). Users generally found the apps easy to use. The GUI received the lowest rating, likely due to their simplistic appearance. We plan to improve upon this in future work. ML capabilities and usefulness were generally rated highly (either 4 or 5) across all three apps. These results indicate that apps created with SmartCS are user-friendly and serve their intended purpose well. This study was conducted through an online user study via open beta testing, where the apps were distributed using Apple's beta testing platform, TestFlight.

\subsection{Qualitative Feedback}
We gathered qualitative feedback from participants of both user studies through open-ended comments and verbal interviews. A primary suggestion from users who created apps was to provide more resources, help files, and video tutorials to guide them through the process. A recurring observation across all study groups was the simplicity and limited features of the user interface. However, this is more of a design choice rather than a technical limitation. Most participants strongly agreed that they appreciated the inclusion of ML for all citizen science use cases. They were impressed with the apps’ ability to detect complex phenomena such as rip currents in real-time. 

\section{Conclusion}
This article presents SmartCS, a platform for building mobile apps with client-side ML-based guidance for citizen science without writing code. The apps created using the platform enhance data collection quality and efficiency through ML-based guidance, even without internet connectivity in the field. The ML guidance also allows the apps to function as educational tools for participants who may not be familiar with the subject of the data being collected. We demonstrate the use of the authoring platform with six example use cases. We also present user studies to illustrate the app creation platform's usability and the effectiveness of the created apps for citizen science applications, highlighting its potential to engage a broader audience in citizen science activities. The feedback suggests areas for improvement, such as offering more resources and enhancing the user interface, but overall, the platform received positive feedback for its usability and the inclusion of ML capabilities.

The current apps enable expert users to utilize the apps without ML support, while non-expert users can benefit from ML assistance for data collection. We plan to facilitate a seamless transition between ML-supported and non-ML modes through in-app guidance in future developments. It is important to note that the ML guidance used in the apps is not infallible and can produce false positive and false negative detections. It would be interesting to see if a collaborative approach between humans and ML can perform even better than just with ML. In such an arrangement, the human will have the option of overriding the ML detection in instances where they are confident and rely on ML detections otherwise. We believe that it is possible to improve overall accuracy and automate further the data collection process. Furthermore, the data collected when humans overrode ML suggestions can be used to improve and refine the ML model further, which we plan to investigate in the future.

\section*{Ethics and Consent}
This research has been approved by the Office of Research Compliance Administration (ORCA), University of California, Santa Cruz and informed consent is obtained from all participants involved in the study.

\section*{Acknowledgments}
This work is partially funded by a grant from the Southeast Coastal Ocean Observing Regional Association (SECOORA), with NOAA financial assistance award number NA20NOS0120220. The statements, findings, conclusions, and recommendations are those of the authors and do not necessarily reflect the views of SECOORA or NOAA.

This project is also funded, in part, by the US Coastal Research Program (USCRP) as administered by the US Army Corps of Engineers® (USACE), Department of Defense. The content of the information provided in this publication does not necessarily reflect the position or the policy of the government, and no official endorsement should be inferred.

This work is also supported by the Google Cloud Research Credits program (award GCP19980904) and AWS Cloud Credit for Research.

We are also grateful for the support and permissions granted by the California State Parks System and the Monterey Bay National Marine Sanctuary to operate our drone for this and related research. We also thank the Santa Cruz Port District, the California State Park Lifeguards and the other beta testing participants 
for their invaluable assistance in evaluating our system and providing feedback to help us improve RipScout.

\appendix

\section{Appendix A: Other Use Cases}
This section briefly describes the other apps created using our platform.

\subsection{TidalNow}
\begin{figure}[ht]
    \centering
    \includegraphics[width=1\linewidth]{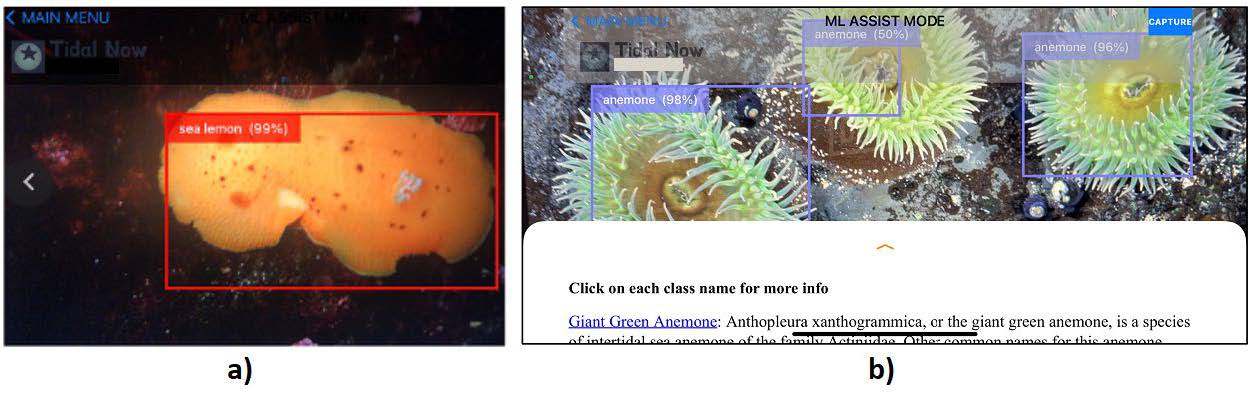}
    \caption{The TidalNow App is shown here. Similar to other apps, this app (a) shows the detected object using a bounding box, (b) the selected template has a built-in pull-up panel that was customized to present additional information about the detected objects.}
    \label{fig:A1}
\end{figure}

There are different dimensions of animal biodiversity (species richness, phyletic richness, and functional diversity) in tide pools. Activities involving observation of wild organisms in the tide pool can provide recreational and learning opportunities \citep{fairchild2018multiple}. TidalNow is a citizen science mobile app developed by a high school student. The app uses a machine learning model to identify different types of saltwater marine species in tide pools. The ML model integrated into the app is trained to detect five different species: giant green anemone, ochre stars, lined chiton, sea lemons, and black turban snails. The ML model for this app was trained using about 600 images for each class. Although apps like iNaturalist or Google Lens can recognize these specimens, they require server-side processing and internet connectivity. However, many tide pools are located on beaches with limited or no internet access. As this app works without internet connectivity, it works perfectly fine in these remote locations (Figure \ref{fig:A1}).

\subsection{Sk.in}
\begin{figure}
    \centering
    \includegraphics[width=1\linewidth]{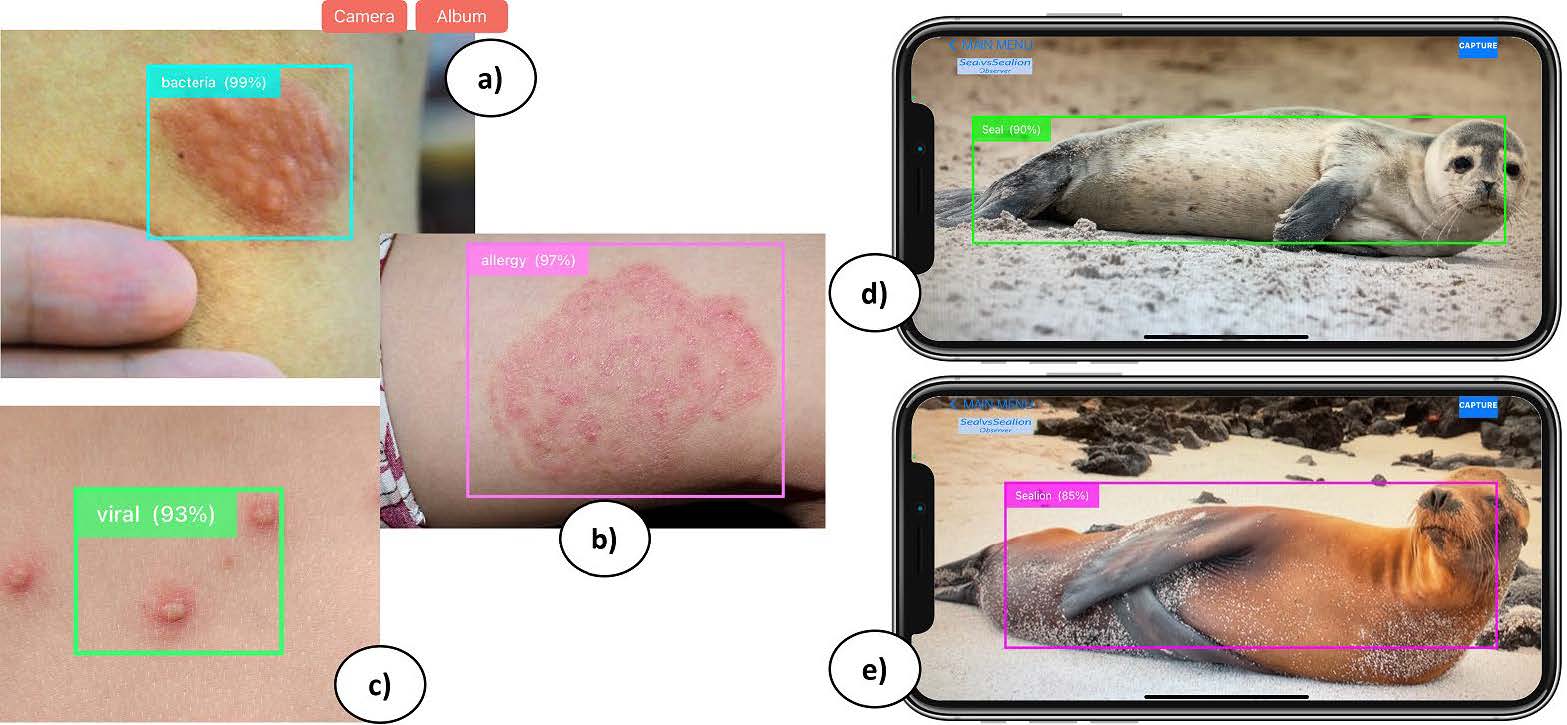}
    \caption{(a)-(c) Demonstrates that the "Sk.in" App can detect bacterial infection, allergy, and viral infection. (d)-(e) Shows example results from the sea lions and seals detection and differentiation app, where detected seals are highlighted using green bounding boxes and sea lions are shown using magenta bounding boxes.}
    \label{fig:A2}
\end{figure}

Skin conditions are more prevalent than other illnesses in all countries worldwide \citep{alenezi2019method}. Some skin diseases can be lethal \citep{allugunti2022machine}. Although the advancement of lasers and photonics based medical technology has made it possible to diagnose skin diseases much more quickly and accurately, the cost of such diagnosis is still very expensive \citep{alenezi2019method}. So, there is a lot of research interest in detecting skin diseases using ML-based computer vision \citep{srinivasu2021classification, shanthi2020automatic, rathod2018diagnosis, alenezi2019method}. Even though some of these recent works demonstrate very accurate skin disease detection using CNN, there are not many works that can do this in real-time on mobile platforms. Sk.in is a mobile application that utilizes ML object detection to categorize dermal conditions as bacterial, fungal, parasitic, viral infections, or allergic reactions in real-time. Sk.in intends to increase the efficiency of diagnosing and treating generalized skin conditions for the public and is designed for everyday use. Developed primarily as an educational tool by a high school student, this app also facilitates data collection on skin infections across a diverse demographic. The ML model can also be trained to detect more skin diseases, such as melanoma and other types of skin cancers \citep{allugunti2022machine}. Figure \ref{fig:A2} (a)-(c) displays the app’s capability of detecting bacterial infection, allergy, and viral infection.

\subsection{Seal vs Sea Lion}
Biodiversity analysis is important for many research groups, such as those with a focus on biological science, aquaculture, marine biology, etc. Researchers may need to collect data about some endangered species; other times, they need data to analyze the biodiversity in some specific area \citep{willi2019identifying, wood2021accuracy}. In this use case, we trained a model with images of sea lions and seals to demonstrate our app’s usability for these types of research projects. Many sea lion species are considered endangered \citep{chilvers2017conservation} and collecting data about them is needed for marine biology research and conservation groups \citep{brown2020california}. However, differentiating between seals and sea lions can be challenging for non-expert participants \citep{wood2021accuracy}. Using our ML-powered app, the participants can detect and differentiate between these two species (Figure 9 (d-e)). With further training data and re-training the model, this app can be modified to detect and differentiate among various sub-species \citep{hann2018obstacles}. The same concept can be applied to create educational and data collection apps about other animal species.

\begin{figure}
    \centering
    \includegraphics[width=1\linewidth]{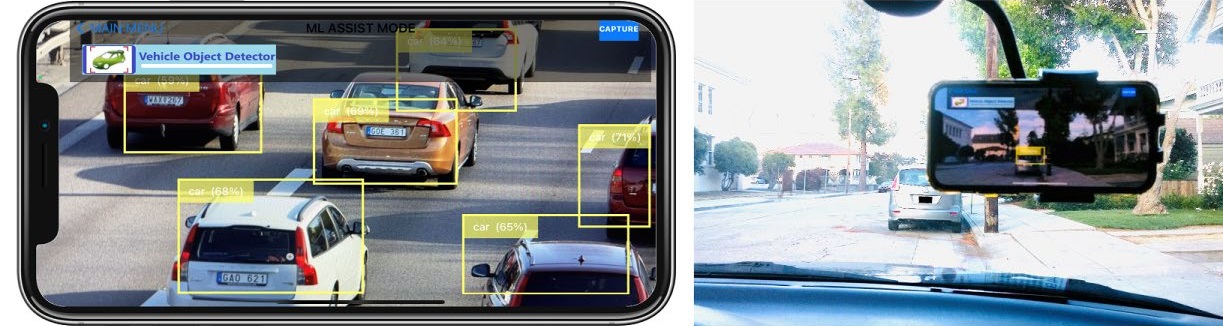}
    \caption{The vehicle object detection app is shown here. The app is used for collecting vehicle data by mounting it on the windshield of a car.}
    \label{fig:A3}
\end{figure}

\subsection{Vehicle Object Detection}
The vehicle detector is a citizen science mobile app for collecting video data about road objects relevant to autonomous vehicle research. It was created as part of an autonomous vehicle research project in collaboration between a university research group and a company from two different countries. The app can be used for road object detection and data collection by mounting the phone next to the windshield of a car. Various standardized datasets exist for autonomous vehicle research, such as Kitti, Waymo, NuScene, etc. \citep{kang2019test}. However, these are collected using arrays of advanced sensors mounted on specialized vehicles \citep{janai2020computer}. This app facilitates small-scale experiments and data collection worldwide, employing a simple and inexpensive mobile setup (Figure 10). Even using the app with multiple mobile devices to collect multiview datasets would be much cheaper than using the traditional autonomous vehicle data collection setup. Therefore, it can be used to collect data using citizen science from different developing countries about various exotic vehicles, such as three-wheelers, rickshaws, etc. These vehicles are not commonly seen in developed countries and, thus, not present in the standard datasets \citep{kato2015open}. 

The other apps developed with the platform include those for detecting plant leaf diseases, identifying beach debris, recognizing types of building architecture, assessing fingernail conditions, counting blood cells, and classifying ultrasound image types.

\section{Appendix B: Summary of Used ML models}

\begin{table}[ht]
\caption{This table presents a summary of the ML models tested and supported on our platform. The information provided here helps the app creators to select a model for training. The faster the inference speed (second column), the app gains better real-time performance. The higher mean average precision (mAP) in the third column represents better precision for detecting objects. Comparing the second and third columns shows that higher precision requires more inference time, leading to slower than real-time performance. The app creator needs to decide about the trade-off between these two. The fourth column shows the approximate final size of the converted trained model, which may impact performance on older devices with lower computational resources.}
\begin{tabular}{|l|l|l|l|}
\hline
\textbf{Model Name} & \textbf{Inference Speed (ms)} & \textbf{mAP for COCO objects} & \textbf{Mobile model size (MB)} \\ \hline
SSD MobileNet v1    & 48                            & 29.1                          & 5                               \\ \hline
SSD MobileNet v2    & 39                            & 28.2                          & 5                               \\ \hline
EfficientDet D0     & 39                            & 33.6                          & 6                               \\ \hline
EfficientDet D1     & 54                            & 38.4                          & 8                               \\ \hline
EfficientDet D2     & 67                            & 41.8                          & 11                              \\ \hline
YOLOv8m             & 32                            & 50.2                          & 49                              \\ \hline
\end{tabular}
\end{table}

The table in this appendix lists the ML models tested on our platform, enabling app creators to select the most appropriate model for training \citep{tensorflow2015-whitepaper}. In our platform's current selection of compatible models for mobile devices released up to 2023, YOLOv8m is the largest recommended model that runs smoothly on a typical consumer mobile device, with a saved weight size of 49 MB. However, even though YOLOv8m shows better performance metrics on benchmarks, we found that EfficientDet D2 offers more stable performance overall during our testing with a few current-generation smartphones (Apple's iPhones and Google’s Pixels). Over time, with the introduction of more powerful mobile devices and larger compatible models, these can be included in our platform without significant modifications.

The number of images needed to train an object detection model can vary widely depending on several factors, such as the complexity of the task, data quality, and model architectures. Decent results can still be achieved even with hundreds of images per class. \cite{shahinfar2020many} suggest an inflection point of around 150-500 images per class, beyond which the earlier sharp performance gains start to level off. However, \cite{bochkovskiy2020yolov4} argue that for optimum accuracy, having at least 2000 different images for each class is desirable. As a rough guideline, for a simple object detection task with a few object classes and relatively consistent object appearances and backgrounds, a few hundred to a few thousand images might suffice, especially if transfer learning is utilized. For more complex tasks or a larger number of object classes, tens or even hundreds of thousands of images might be needed.

The number of classes supported for training depends on the model type and architecture. For instance, according to official documentation, EfficientDet can support up to 999 classes \citep{tan2020efficientdet}, whereas YOLOv8 (Ultralytics, 2023) does not have a defined hard limit for the number of classes. However, the choice of model determines the number of classes; it is not a limitation of our platform, SmartCS, since we can incorporate newer versions of models that support more classes.

Although transfer learning cannot be performed within the app due to technological limitations and the resource constraints of mobile devices, which render model training infeasible, models can be updated periodically with newly collected data by retraining them using transfer learning on more powerful machines or cloud servers. Additionally, we provide pre-trained models derived from well-known public image datasets, such as MS-COCO, to serve as a starting point for training on new datasets via transfer learning.

\bibliography{references}

\end{document}